 \documentclass[preprint2]{aastex}

\begin{document}


\title{Dynamical Rotational Instability at
Low $T/W$}

\author{Joan M. Centrella\altaffilmark{1}, Kimberly
C. B. New\altaffilmark{2,3}, Lisa L. Lowe\altaffilmark{1}, and
J. David Brown\altaffilmark{4}}

\altaffiltext{1}{Department of Physics, 
Drexel University, Philadelphia, PA 19104}
\altaffiltext{2}{X-2, MS B-220,
Los Alamos National Laboratory,
Los Alamos, NM 87545}
\altaffiltext{3}{The initial stage 
of this work was conducted while New
was at the Department of Physics, Drexel University.}
\altaffiltext{4}{Department of Physics, North Carolina State University,
Raleigh, NC 27695}

\begin{abstract}
Dynamical instability is shown to occur in differentially rotating
polytropes with $N = 3.33$ and $T/|W| \gtrsim 0.14$.  This instability
has a strong $m=1$ mode, although the $m=2, 3,$ and $4$ modes
also appear.  Such instability may allow a centrifugally-hung core to
begin collapsing to neutron star densities on a dynamical timescale.
The gravitational radiation emitted by such unstable cores
may be detectable with advanced
ground-based detectors, such as LIGO II.
If the instability occurs in a supermassive star, it may produce gravitational
radiation detectable by the space-based detector LISA.
\end{abstract}


\section{Introduction}

Rotational instabilities are potentially important in the evolution
of massive stellar cores.  For example, the core of a
massive star that has been prevented from 
collapsing to neutron star densities by centrifugal forces
may rotate rapidly enough for rotational instabilities to 
develop \citep{HEH98,HEH99}. 
Massive cores spun up by accretion from a
binary companion \citep{wagoner84} and the remnants of compact binary
mergers \citep{RS92,ZCM94} may also reach fast enough 
rotation rates to become
unstable. Rotational instabilities 
may produce
detectable gravitational
radiation \citep{schutz89,thorne96}.  If enough
angular momentum is shed, full collapse to
a neutron star or black hole may occur.

We focus here on global rotational instabilities that arise in
fluids from growing azimuthal modes $e^{i m \phi}$
\citep{tassoul}.  Dynamical 
instabilities are driven by hydrodynamics and gravity, and grow
on the order of the dynamical timescales of the system.
In objects with
Maclaurin-like rotation laws,  dynamical rotational
instabilities are generally expected to arise 
 at fairly high values of 
 $\beta \equiv T/|W| \gtrsim 0.27$ in Newtonian gravity
(see New, Centrella, \& Tohline and references therein), and 
 $\beta  \gtrsim 0.25 - 0.26$ in general relativity
\citep{SBS00}.  Here, $T$ is the rotational
kinetic energy, and $W$ is the gravitational potential energy.
 Dynamical instability may set in at lower values of
$\beta$ for thick, self-gravitating disks
\citep{PDD,WTH} and in tori \citep{TH}.  Secular instabilities develop
on longer timescales, and arise from dissipative processes such as
viscosity and gravitational radiation reaction.  They are expected
to set in for relatively low values of $\beta \gtrsim 0.14$, although
the rotation law, polytropic index, and general relativity can
affect the value of the instability limit \citep{ITDPY95,SF98}.

Studies of scenarios for the
formation of centrifugally-hung
cores 
have generally concluded that these objects will have 
$\beta < 0.27$ \citep{tohline84, EM85, ME85}. 
Direct numerical
simulations of axisymmetric 
stellar collapse by \citet{ZM97} also
indicate that, for the cases studied,
 it is difficult for collapsing cores to reach
and exceed $\beta \sim 0.27$.   Overall, it is
generally assumed that centrifugally-hung cores will evolve secularly,
as in the ``fizzler'' scenario \citep{HEH98,HEH99}.

In this {\it Letter}, we report on new 3-D numerical simulations
of rotating stellar cores that exhibit dynamical rotational
instability
at relatively low values of $\beta \gtrsim 0.14$. 
These models are computed using Newtonian gravity with no
backreaction; this is a reasonable approximation for a stellar core
of mass $M \sim 1.4 M_{\odot}$ hung up at a radius
$R \gtrsim 100$km, so that $(GM/Rc^2) \lesssim 0.02$.
Our results show that compact objects can become dynamically
unstable in a region of parameter space previously thought to be
solely the province of secular instabilities.

\section{Numerical Simulations}

Rotating stellar cores are initially assumed to be axisymmetric
equilibrium polytropes with equation of state $P = K\rho^{\Gamma}
= K\rho^{1+1/N}$, with $\Gamma = 1.3$ ($N = 3.33$).
The angular velocity distribution is given by the
so-called $j-$constant rotation law,
$\Omega^2 = j_o^2/(d^2 + \varpi^2)^2$,
where $\varpi$ is the cylindrical radius and $d$ is an arbitrary
constant \citep{hachisu86}. As $d \rightarrow 0$, the specific angular
momentum approaches the constant value $j_o$. Here,
we use $d = 0.2$.  The models are computed
on a uniformly spaced $(\varpi,z)$ grid 
using the 
self-consistent field method of
\citet{hachisu86}.  Varying the axis ratio yields
equilibria
with different values of $\beta$; see \citet{NCT00} for details.
Figure~\ref{init-mods} shows density contours in the $x-z$ plane
for the 4 models that we evolved here, $\beta = 0.090, 0.12, 0.14,$
and $0.18$.
Notice that, with the exception of the $\beta = 0.090$ model, the
density maxima are toroidal; similar structures were 
used as initial data in the
collapse models of \citet{RMR98}.

These initial models were introduced into 2 different 3-D
hydrocodes.  The first of these is the {\cal L} code discussed in
\citet{NCT00}, with a cylindrical $(\varpi,z,\phi)$ grid
of  $64 \times 64 \times 128$ zones. The
second is the PPM code discussed by \citet{DB00},
with a Cartesian $(x,y,z)$ grid of $128 \times 128 \times 128$ zones.
Both codes have the same resolution along the $x$ and $z-$axes,
with the equatorial radius $R_{\rm E}$ of the 
initial model extending out to zone 48 in $x$-direction. The
cylindrical code imposes equatorial plane symmetry, whereas the Cartesian
code imposes no symmetries. Random
perturbations of 1\% were imposed on the densities, and the
models were evolved forward in time.

\section{Results}

The model with initial $\beta = 0.14$ exhibits
a dynamical instability that is very similar
in simulations performed with both the
cylindrical and Cartesian hydrocodes. These runs were stopped when the
loss of mass from the edge of the grid became significant. 
Figure~\ref{contours-cyl} shows density contours in the equatorial
plane from the run with the cylindrical code. 
The green (lightest greyscale) contour
in the first frame delineates the toroidal
density maximum seen in Fig.~\ref{init-mods}. As time proceeds,
this torus ``pinches off'' to leave a single high density region,
indicative of an $m=1$ mode.  At late times, the appearance of higher
density contours signals that the dense region is starting to
collapse.
Time is measured in units of
$t_{\rm D}=
 (R_{\rm E}^3/GM)^{1/2}$, which is the dynamical time for a {\em sphere}
of mass $M$ and radius  $R_{\rm E}$.  For $R_{\rm E} \sim
200$ km, $t_{\rm D} \sim 6$ ms.
3-D animations of the evolution of this model, using
the same contour levels as in Figure~\ref{contours-cyl} and looking
down the $z-$axis towards the equatorial plane, are available for both
the cylindrical and Cartesian runs.  Additional animations, in which
the model is viewed from the side, are also available for the
cylindrical and Cartesian runs. [Note to editors: Here, the words
``cylindrical'' and
``Cartesian'' should link to the animations in each case.]  

We can quantify the instability
by examining various Fourier components
in the density distribution.  We examine the
density in a ring of fixed $\varpi$ and $z$ 
using a complex azimuthal Fourier decomposition,
where the amplitudes $C_m$
of the various components $m$ are defined by 
$C_m(\varpi,z) = (1/2 \pi) \int_0^{2 \pi} \rho(\varpi,
        z,\varphi) e^{-im\varphi} d\varphi$ \citep{TDM}.
The normalized amplitude is then
$A_m = C_m/C_0$,
where $C_0(\varpi,z) = \bar{\rho}(\varpi,z)$ is the 
mean density in the ring, and the phase angle of 
the $m^{\rm th}$ component is
$\phi_m(\varpi,z) = \tan^{-1}\left [ {\rm Im}(A_m)/{\rm Re}(A_m)
\right ]$.
We write the phase angle
$\phi_m = \sigma_m t$, where $\sigma_m = d\phi/dt$ is the
eigenfrequency and $W_m = \sigma_m/m$ is the pattern speed
of the $m^{\rm th}$ mode.

Figures~\ref{modes-CYL} and~\ref{modes-CRT} 
show the growth of the amplitudes $|A_m|$ for the first four Fourier
components in the runs with $T/|W| = 0.14$ on the cylindrical
and Cartesian hydrocodes, respectively.
For the cylindrical code, these amplitudes were calculated in the
equatorial plane in a ring of width $\Delta \varpi = 1/48$ at
radius $\varpi = 0.32$; see \citet{NCT00} for details.
For the Cartesian code, the amplitudes were computed on a circle
of radius $\varpi=0.32$ using a nonuniform discretization, which
avoids grid boundaries, and a linear interpolation of density.

Similar results were found in rings at other values of $\varpi$.
The evolution in both runs
is dominated by an exponentially growing
$m=1$ mode (thick solid line).
Once this mode is well into its exponential growth phase, an
$m=2$ mode also develops, followed at later times by
$m=3$ and $m=4$ modes.  (The constant amplitude $m=4$ signal in the
Cartesian run is due to the geometry of the numerical grid.)
Since these modes grow rapidly on dynamical timescales, they signal
dynamical instability. 
Numerical values for the growth rates, eigenfrequencies, and pattern
speeds are shown in Table~\ref{modes} for $m=1$ and $m=2$.  Notice
that the cylindrical and Cartesian runs yield similar results and that
the pattern speeds $W_1 \sim W_2$, indicating that the
$m=2$ mode is a harmonic of the $m=1$ mode.  Longer runs with larger
grids and higher resolution are needed to obtain reliable values for
the $m=3$ and $m=4$ modes.

The model with $\beta = 0.18$ also exhibits dynamical instability.
For the evolutions run with both codes,
the $m=1$ mode grew at very close to the same rate,
starting at about the same time; see Table~\ref{modes}.
In the run on the cylindrical code, the modes $m=1,2,3,$ and $4$ 
appear sequentially
in time and $W_1 \sim W_2$, as in the model with $\beta = 0.14$.
In the run on the Cartesian code, the $m=1$ and $m=2$ modes
both grew at about
the same rate, starting at about the same time;  however, $W_1 \neq
W_2$.
At this stage, we do not understand this difference in the behavior
of the $m=2$ mode in this case.  We plan to investigate it using
higher resolution simulations.
The $\beta = 0.12$ model was run for $\sim 40 t_{\rm D}$, at
which time the $m=1$ mode was just beginning to grow;
longer runs with higher resolution are needed to confirm
the development of instability in this case.
The model with $\beta = 0.09$ was
run for $\sim 35 t_{\rm D}^{-1}$, and showed no growing modes.  In the
event that we can confirm instability for $\beta = 0.12$, we will
return to the $\beta = 0.09$ case and run it further.

Previous studies of the $m=2$ bar-mode instability using the
cylindrical hydrocode
showed that significant motion of the system center of mass could
develop
at late times, resulting in a spurious $m=1$ signal \citep{NCT00}.
For the simulations reported here, we monitored the position of the
overall center of mass and verified that it underwent no systematic
motion during the development of the $m=1$ mode, in both the
cylindrical and Cartesian codes. When the
runs with $T/|W| = 0.14$ and $T/|W| = 0.18$ were stopped,
some small, spurious center of mass motion ($<$ one zone)
had developed in both codes.  This effect is
possibly related to the loss of mass from the
grid as the outer regions expand; runs with
larger grids are needed to determine the late time behavior of these
models. 
Previous studies of the bar instability using the Cartesian code
\citep{DB00} showed that angular momentum losses due to numerical
inaccuracies can be signficant.  For the $\beta=0.14$ and $\beta=0.18$
simulations with the Cartesian code, the artificial angular momentum
losses amounted to less than $0.4\%$ and $1.5\%$, respectively.

\section{Discussion}
Our results demonstrate that dynamical instability can occur in
differentially rotating polytropes with low values of $\beta \gtrsim
0.14$, and that this instability has a strong, $m=1$
character.  Figure~\ref{init-mods} shows that the unstable
axisymmetric equilibria each have a torus of dense material centered
on the rotation axis within the model.  This suggests that this
instability may be related to the ones found by \citet{WTH} and
\citet{TH} in their studies of polytropic tori with $N=3/2$.
We note that \citet{PDD} also found a dominant $m=1$ instability that
set in at a relatively low $\beta \gtrsim 0.20$, in a 
centrally condensed model (without an off-center density maximum).
Their model was an
$N=3/2$ polytrope with an $n'=2$ rotation law.  The $n'=2$ rotation law
produces configurations with a strong concentration of angular momentum
in the outer regions of the model.
We plan to
carry out more detailed studies to investigate the character of the
unstable modes seen in our simulations
and their properties for various values of $N$ and the
$j$-constant rotation law parameter $d$.

If such instability occurs in a centrifugally-hung core, collapse to
neutron star densities may result.  Our simulations do show that the
density is increasing at the end of the unstable runs.  Further
studies are needed to see how dense the remnant actually becomes, and
whether the $m=1$ mode will result in the dense region moving at a
velocity comparable to those of actual neutron stars \citep{popov}. 

Dynamical instability will also produce gravitational radiation.
Although longer runs are needed to obtain the full gravitational
waveforms, we can estimate the properties of the gravitational wave
emission from the initial stages of the instability.  For a massive
stellar core with 
$M \sim 1.4 M_{\odot}$ and $R_{\rm E} \sim 200$km, the peak amplitude will
occur at a frequency of roughly $f \sim 200$Hz.  This peak amplitude
will be $h \sim 10^{-24} r_{20}^{-1}$ for $\beta = 0.14$, and
$h \sim 10^{-23} r_{20}^{-1}$ for $\beta = 0.18$.  Here, $r_{20}$ is
the distance to the source in units of 20 Mpc.
Emission from such unstable cores may be detectable with advanced
ground-based interferometers, such as LIGO-II.
An even more optimistic
scenario for detectable gravitational radiation occurs if these
instabilities arise in supermassive stars\footnote{
We note that the appropriate equation of state for supermassive stars is
the $n=3$ polytropic equation of state.} \citep{NS}.
For example, if the instability occurs when a supermassive star
contracts to the
point that $(GM/Rc^2) \sim 1/15$,
an approximate value for uniformly
rotating stars (Baumgarte \& Shapiro 1999), we estimate the
frequency of the gravitational radiation to be $f \sim 3.5 \times
10^{-3}$Hz, with amplitude $h \sim 10^{-18}r_{20}^{-1}$ for $\beta =
0.14$ and $h \sim 10^{-17}r_{20}^{-1}$ for $\beta = 0.18$.
Such signals would be easily
detectable by the space-based LISA detector.

\acknowledgments
We thank K. Thorne and J. Tohline for stimulating discussions, and
E. Mamikonyan for assistance with the animations.
This work was supported in part by NSF grants
PHY-9722109, PHY-0070892, 
NSF cooperative agreement ACI-9619020 through computing
resources provided by the National Partnership for Advanced
Computational Infrastructure (NPACI) at the San Diego Supercomputer
Center (SDSC), and the North
Carolina Supercomputing Center (NCSC). 
A portion of this work was performed under the
auspices
of the U.S. Department of Energy by Los Alamos National Laboratory under
contract
W-7405-ENG-36.
The cylindrical simulations were run on the T3E at SDSC, and
the Cartesian models on the T90 at NCSC.



\figcaption[init-mods.eps]{Density contours are shown in the
$x-z$ plane for 4 models with $d=0.2$ and $N=3.33$. 
The initial maximum density is normalized to unity, and the contours
are at levels of 0.9, 0.1, 0.01, and 0.001\label{init-mods}}

\figcaption[contours-cyl.eps]{2-D density contours in the equatorial plane
are shown for the model with $\beta = 0.14$ run on the cylindrical
code.
The contour levels are 0.01 (purple), 0.1 (blue), 0.9 (green), 2 (yellow),
and 4 (off-white) times the maximum density at the
initial time, which is normalized to unity.  In the greyscale version
of this figure,
the density decreases as the darkness of the shading increases.
\label{contours-cyl}}

\figcaption[modes-CYL.eps]{The growth of the amplitudes $|A_m|$ for
$m=1$ (thick solid line), $m=2$ (thin solid line),
$m=3$ (dashed line), and $m=4$ (dotted line)
is shown for the model with initial $T/|W| = 0.14$ run on the
cylindrical hydrocode.  These amplitudes were
calculated in the equatorial plane for a ring with
radius $\varpi = 0.32$.  \label{modes-CYL}}

\figcaption[modes-CRT.eps]{Same as Figure~\protect\ref{modes-CYL},
except that the model was evolved on the Cartesian hydrocode.
 \label{modes-CRT}}


\begin{table}[p]
\begin{center}
\begin{tabular}{cccccccc}
\tableline
code & $\beta$ & $d \ln |A_1|/dt$  & $d \ln |A_2|/dt$  & $\sigma_1$ &
 $\sigma_2$ & $W_1$ & $W_2$ \\
&  & [$t_{\rm D}^{-1}$] & [$t_{\rm D}^{-1}$] & [$t_{\rm D}^{-1}$] &
[$t_{\rm D}^{-1}$] & [$t_{\rm D}^{-1}$] & [$t_{\rm D}^{-1}$] \\
\tableline
cyl  & 0.14 & 0.40 & 0.92 & 3.6 & 7.2 & 3.6 & 3.6 \\
cart & 0.14 & 0.35 & 0.90 & 3.5 & 7.2 & 3.5 & 3.6 \\
cyl  & 0.18 & 0.99 & 1.8  & 3.3 & 6.6 & 3.3 & 3.3 \\
cart & 0.18 & 0.98 & 1.1  & 3.2 & 2.4 & 3.2 & 1.2 \\
\tableline
\end{tabular}
\end{center}
\caption{The growth rates, eigenfrequencies, and pattern speeds for
the $m=1$ and $m=2$ modes are given
in a ring at radius $\varpi = 0.32$ in the
equatorial plane.}
\label{modes}
\end{table}

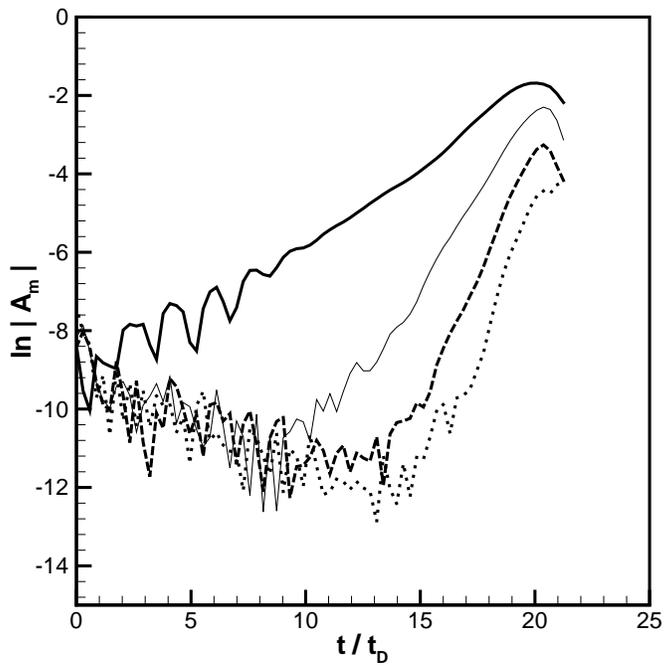
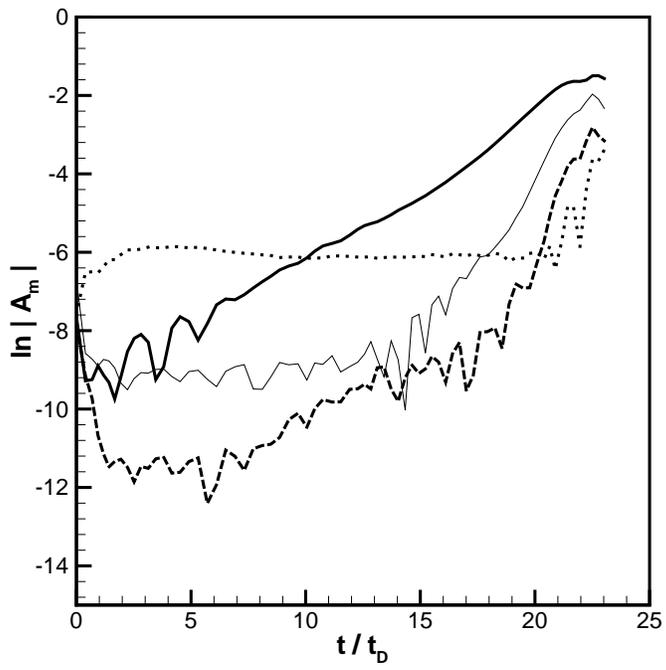
\end{document}